\documentclass{jfpda2014}

\usepackage{amsmath, amssymb,algorithmic,algorithm}
\usepackage{natbib, subfigure}
\usepackage{color}

\newcommand{\bfN}{\boldsymbol N}
\newcommand{\Exp}{\mathbb E}

\newcommand{\boldeta}{\boldsymbol \nu}
\newcommand{\bfc}{\boldsymbol c}

\newcommand{\calF}{\mathcal F}
\newcommand{\hatQ}{\hat Q}
\newcommand{\tildeQ}{\tilde Q}

\newcommand{\se}{\mathrm{se}}

\newcommand{\state}{\boldsymbol n}

\newcommand{\action}{\boldsymbol a}
\newcommand{\actionSet}{\boldsymbol A}

\newcommand{\rew}{\boldsymbol c}

\newcommand{\figref}[1]{Figure~\ref{#1}}
\newcommand{\eg}{e.g.,~}
\newcommand{\ie}{i.e.,~}
\newcommand{\laci}{{\it lacI}}
\newcommand{\tetr}{{\it tetR}}
 \newcommand{\incfig}[1]{#1}
\newcommand{\argmin}{\mathop{\text{argmin}}}
\title{Toggling a Genetic Switch Using Reinforcement Learning}
\author{Aivar Sootla\inst{1}, Natalja Strelkowa\inst{2}, Mauricio Barahona\inst{3}, Damien Ernst\inst{4}, Guy-Bart Stan\inst{1}}
\institute{Centre for Synthetic Biology and Innovation and the Department of Bioengineering, Imperial College London, UK {\tt\small \{a.sootla, g.stan\}@imperial.ac.uk}
\and
Boehringer-Ingelheim Pharma GmbH \& Co KG., Germany {\tt\small natalja.strelkowa@boehringer-ingelheim.com}
\and 
Department of Mathematics, Imperial College London, UK {\tt\small m.barahona@imperial.ac.uk}
\and 
Montefiore Institute, University of Li\`ege, Belgium {\tt\small dernst@ulg.ac.be}.}

\begin{document}

\maketitle
\begin{abstract}
In this paper, we consider the problem of optimal exogenous control of gene regulatory networks. Our approach consists in adapting an established reinforcement learning algorithm called the fitted Q iteration. This algorithm infers the control law directly from the measurements of the system's response to external control inputs without the use of a mathematical model of the system. The measurement data set can either be collected from wet-lab experiments or artificially created by computer simulations of dynamical models of the system. The algorithm is applicable to a wide range of biological systems due to its ability to deal with nonlinear and stochastic system dynamics. To illustrate the application of the algorithm to a gene regulatory network, the regulation of the toggle switch system is considered. The control objective of this problem is to drive the concentrations of two specific proteins to a target region in the state space. 
\end{abstract}

\section{Introduction}
Synthetic biology aims at the (re-)design of biological functions in living organisms for their use in various applications such as bioengineering, bioremediation and energy (\cite{Purnick:2009}). This is typically realised via the insertion of foreign genes inside a host cell (e.g., a bacterium {\it E. coli}). The expression of the foreign genes inside the host cells imposes \emph{de facto} a burden on the native processes of the host cells. A high burden induces severe intracellular perturbations and can decrease cellular growth rate. This in turn disrupts the intended behaviour of synthetic biology gene networks (\cite{Tan:2009}). Hence, it is highly desirable to develop means for controlling gene networks so as to efficiently enable the designed behaviour while simultaneously minimising the burden induced by this behaviour on the host cells.

The current biotechnology state-of-the-art allows us to quantitatively measure and interact with gene regulatory networks. Quantitative \emph{in vivo} estimates of gene networks' states (outputs) can be obtained via fluorescent markers (\cite{Xie06,Bennett09}) (\eg green fluorescent protein, GFP or red fluorescent protein, mCherry). A typical input is a targeted induction of the gene expression, which can be achieved by, e.g., conditional gene knock outs (\cite{Ivanova06, Liu07}), heat shocks (\cite{Mettetal08}) or monochromatic light pulses (\cite{Sato02, Levskaya09}). This means that feedback control is technologically feasible \emph{in vivo}. The objective of the control method can be minimal time control (\ie driving the system as fast as possible to a target region in the state-space), minimal burden control (minimal expression of heterologous proteins), or a trade-off between the two, as considered in this paper. The control method must reach the objective, while maintaining the designed functions of a
synthetic gene regulatory network.


Some control problems in gene regulatory networks were successfully addressed (\cite{Menolascina:2011,uhlendorf2012long,Milias2011silico}). In all those papers, the authors used classical control methods, which infer the control law (or the control policy) based on a mathematical model of the system. One of the bottlenecks of these approaches is the modelling part, which for large gene regulatory networks is an extremely hard and lengthy process. Moreover, there are other challenges such as stochasticity. Stochasticity is expressed in the form of the intrinsic and extrinsic noise during gene expression (\cite{Swain02}). Transcription and translation processes typically involve a few randomly interacting molecules, thus adding thermodynamic stochasticity to biochemical interactions. 

\begin{figure}[t]
\centering
\includegraphics[width=0.5\columnwidth]{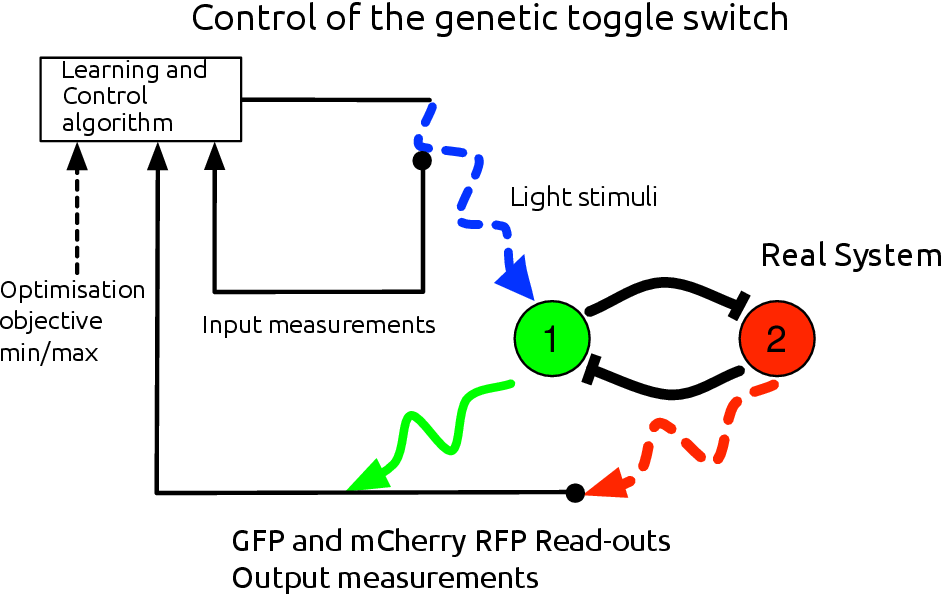}
\caption{A schematic depiction of the exogenously controlled genetic toggle switch. The green circle represents the \laci~gene and the red circle represents the \tetr~gene. The arrows with flat ends represent repression of one gene by another. In the steady-state only one of the genes can be upregulated (or switched on). The goal is to toggle one of the genes, \ie drive this gene from its downregulated mode to its upregulated one.
} \label{fig:ts}
\end{figure}

The problems with modelling and stochasticity point towards the use of reinforcement learning methods (\cite{sutton1998, busoniu2010reinforcement}), which infer the control policy based solely on interactions with the real system. These methods do not require a physical model. Moreover, very few assumptions on the structure of the controlled system are made. However, the major advantage of the reinforcement learning methods is to some extent their drawback. Indeed, these methods require interactions with the real system, which implies numerous costly and lengthy wet-lab experiments. A solution would be a reinforcement learning method, which learns the policy using a single experiment. For systems relevant to this paper, however, such a method will not be efficient. Indeed, a control policy, which tries to learn and control such systems in a single experiment, is generally not better than a random control policy (\cite{castronovo2012learning}). In order to address these concerns, a hybrid approach is proposed.
First, an initial control policy is computed using past experimental data and/or a mathematical model. After that the control policy is updated during the experiment using reinforcement learning methods. This approach will be applied to the regulation of the toggle switch system schematically depicted in \figref{fig:ts}. The control objective of this problem is to drive the concentrations of two specific proteins to a target set in the state space and remain in this set.

The initial policy is obtained by the Fitted $Q$ Iteration algorithm (\cite{Ernst2005}). The algorithm requires only one-step system transitions to infer the control policy. A one-step system transition is a triplet $\{\state, \action, \state^+\}$, where $\state^+$ denotes a successor state of the system in state $\state$ subjected to input $\action$. Fitted $Q$ Iteration can also handle nonlinear and stochastic systems and it is sample efficient. One-step transitions can be obtained by simulating the mathematical model of the system or using past experimental data. Afterwards the policy is updated by mixing the online measurements with past observations. The Exploration/Exploitation trade-off is addressed using an $\varepsilon$-greedy policy.

This paper is organised as follows. Mathematical preliminaries are described in Section~\ref{s:prel}. In order to make the paper self contained, the fitted $Q$ algorithm is sketched and different aspects of modelling in gene regulatory networks are discussed. The problem of controlling the toggle switch is formulated and discussed in detail in Section~\ref{s:prob}. Finally, the simulation results are presented in Section~\ref{s:res}. Reference trajectory tracking for the generalised repressilator system is the subject of our previous publication (\cite{sootla2013tracking}).

\section{Preliminaries \label{s:prel}}
\subsection{Modelling in Biology}
The following approach to chemical reaction modelling is described in detail in (\cite{gillespie1977exact}).  Consider a well-stirred system of $k$ species in a constant volume $\Omega$ and a thermal equilibrium. Assume the species are interacting through $m$ reactions. Let $N^i(t)$ be the number of molecules of species $i$ and $\nu_{ij}(t)$ be the change in the molecular concentration of species $i$ at time $t$ if the reaction $j$ occurs. The bold symbols will be used to denote vectors, e.g., $\state$ stands for the vector with elements $N^i$. Finally, let $a_j(\state) d t$ be the probability of reaction $j$ occurring in the next infinitesimal interval $[t,t+d t]$, if the number of molecules at time $t$, $\bfN(t)$, is equal to $\state$. The functions $a_j(\cdot)$ are called propensity functions. At the cellular level chemical reactions depend on thermodynamical principles, since molecules must collide before a reaction can start. Therefore chemical reactions inside living organisms are modelled using stochastic calculus. The time evolution of the concentration of species can be modelled by a Markov stochastic process,
for which:
\begin{equation}
\frac{\partial \Pr(\state, t | \state_0, t_0)}{\partial t} = \sum\limits_{j =1}^m a_j(\state - \boldeta_j) \Pr(\state-\boldeta_j, t| \state_0, t_0 ) - 
a_j(\state) \Pr(\state, t| \state_0, t_0) \label{eq:CME}
\end{equation}
where the probability $\Pr(\state, t| \state_0, t_0)$ stands for $\Pr(\bfN(t) = \state| \bfN(t_0) = \state_0)$. This equation is called the \emph{Chemical Master Equation}. The propensity functions $a_j$ depend also on the volume $\Omega$. It can be shown that for large volumes $\Omega$ the CME~\eqref{eq:CME} becomes a deterministic equation 
\begin{equation}\label{eq:det}
\frac{d \state(t)}{d t} = \sum\limits_{j =1}^m \boldeta_j \tilde a_j(\state(t)),
\end{equation}
where the propensities $\tilde a_j$ are independent of the volume $\Omega$. For small volumes $\Omega$, the stochastic model~\eqref{eq:CME} describes better the behaviour of the cells than the deterministic model \eqref{eq:det}. Hence, in synthetic biology setting using a stochastic model is preferable. Nevertheless the deterministic model can be still useful for small volumes in order to provide some idea of the system behaviour, since stochastic models are harder to simulate and analyse.
\subsection{Formulation of the Optimal Control Problem}
Consider a deterministic discrete-time dynamical system
\begin{equation}
 \begin{aligned}
  \state_{t+1} = f(\state_t, \action_t)
 \end{aligned}\label{eq:sys}
\end{equation}
where $\action_t$ is the control input at time $t$, which belongs to a compact set $\actionSet$ for every $t$. In the stochastic case, Markov decision processes (MDPs) are typically employed, for which
\[
 \Pr\left(\state_{t+1} \in \bfN_{t+1}\Bigl| \{\state_k\}_{k=0}^{t}, \{\action_k\}_{k=0}^{t}\right) = 
  \Pr\left(\state_{t+1} \in \bfN_{t+1} \Bigl| \state_t, \action_t\right).\]
The above relationship means that the probability of the state $\state_{t+1}$ belonging to the set $\bfN_{t+1}$ does not depend on the entire history of the realisation of the states $\{\state_k\}_{k=0}^{t}$ and control signals $\{\action_k\}_{k=0}^{t}$, but depends only on the current values $\state_t$ and $\action_t$. Under the above Markovian assumption, dynamical stochastic systems can be modelled as
\[
  \Pr\left(\state_{t+1}\in \bfN_{t+1} \Bigl| \state_t, \action_t\right) = \int\limits_{\bfN_{t+1}} f(\state_t, \action_t, x ) \,d x,
\]
or in a compact form
\[
 \begin{aligned}
  \state_{t+1} \sim f(\state_t, \action_t, \cdot). 
 \end{aligned}
\]
Here, we slightly abuse the notation by using again the symbol $f$ as in the deterministic system~\eqref{eq:sys}. This is done, in order to signify that these functions describe the dynamics of the system whether it is stochastic or deterministic.

In both cases, consider an optimal control problem, which is defined through the minimisation of an infinite sum of discounted costs $\bfc(\state, \action)$. In the deterministic case the problem is defined as
\[
V(\state_t) = \min_{\pi(\cdot):~\pi(\state_i) = \action_i}\sum_{i=t}^{\infty} \gamma^{i-t} \bfc(\state_i, \action_i)
\]
and in the stochastic case as
\[
V(\state_t)= 
\min_{\pi(\cdot):~\pi(\state_i) = \action_i} \lim\limits_{K\rightarrow\infty}\Exp_{\state_{t+1} \sim f(\state_t, \action_t, \cdot)}\sum_{i=t}^{K} \gamma^{i-t}\bfc(\state_t, \action_t)
\]
where $V(\state_t)$ is called the value function and $\pi(\cdot)$ is a mapping from $\state$ to $\action$, which is called the control policy. The cost function $\bfc$ specifies the objective of the control problem, which in our case is driving the system to a specific region in the state-space. In our setting, the control policy should be inferred based only on realisations of one-step system transitions $\{\state_l, \action_l, \state_l^+\}$, where $\state_l^+$ is a successor state of the system in the state $\state_l$ and subjected to the input $\action_l$ (in the deterministic case, if the function $f(\cdot,\cdot)$ is known $\state_l^+$ is equal to $f(\state_l,\action_l)$). For the purpose of this paper, the function $\bfc(\cdot,\cdot)$ is assumed to be known in advance.

\subsection{Fitted $Q$ Iteration}
A central object of the fitted $Q$ algorithm is the $Q$ function, which is introduced as follows:
\[
Q(\state_t, \action_t) = \bfc(\state_t, \action_t) +  \min_{\pi(\cdot)}\sum_{i=t}^{\infty} \gamma^{i-t} c(\state_i, \pi(\state_i))
\]
Once a $Q$ function is computed, the optimal feedback control policy is given as:
\[
\pi^{\ast}(\state) = \argmin_{\action \in \actionSet} Q(\state, \action)
\]
Under certain conditions, the $Q$ function can be obtained as the unique solution of the following iterative procedure:
\begin{equation} \label{eq:bellman}
\begin{aligned}
Q_k(\state, \action) = \bfc(\state, \action) + \gamma \min_{\action'\in \actionSet}Q_{k-1}(f(\state,\action),\action')
\end{aligned}
\end{equation}
where $Q_0$ is equal to $\bfc$. However,~\eqref{eq:bellman} is hard to solve in general, especially if only the triplets $\calF$ are given. Therefore an approximation $\hatQ$ of the $Q$ function is computed using an iterative procedure. Let $\hatQ_0 = \bfc$  and for every $(\state_l,\action_l,\state_l^+)$ in $\calF$ compute:
\[\hatQ_{1}(\state_l, \action_l) = \bfc(\state_l, \action_l) + \gamma \min_{\action\in \actionSet}\hatQ_{0}(\state_l^+, \action)\]
This expression gives $\hatQ_1$ only for $\state_l$, $\action_l$ in $\calF$, while the entire function $\hatQ_1(\cdot,\cdot)$ is estimated by a regression algorithm (\eg EXTRA Trees by \cite{Geurts:2006}). This can be generalised to an iterative procedure, which can be used to obtain a near-optimal control policy as outlined in Algorithm~\ref{alg:fqi}. The stopping criterion can be simply the maximum number of iterations $N_{\rm it}$, which is chosen such that the number $\gamma^{N_{\rm it}}$ is sufficiently small and the values $\hatQ_k(\state_l, \action_l)$ are not modified significantly for $k$ larger than $N_{\rm it}$. Other criteria are described in~ (\cite{Ernst2005}). 
Note that Algorithm~\ref{alg:fqi} can be extended to handle the stochastic case as well (\cite{Ernst2005}).
\begin{algorithm}[t]
\caption{Fitted $Q$ iteration algorithm~\label{alg:fqi}}
{\bf Inputs:} Set of triplets $\calF = \{\state_l, \action_l, \state_l^+\}_{l=1}^{\#\calF}$, stopping criterion, cost function $\bfc(\cdot,\cdot)$\\
{\bf Outputs:} Policy $\hat\pi^{\ast}(\state)$
\begin{algorithmic}
\STATE $k \leftarrow 0$
\STATE $\hatQ_0(\cdot,\cdot)\leftarrow \bfc(\cdot,\cdot)$
\REPEAT
\STATE $k \leftarrow k+1$
\STATE In order to obtain the values of $\hatQ_k(\cdot, \cdot)$ for all $\{\state_l, \action_l\}$ in $\calF$ compute:
\begin{equation}
\label{eq:fitQ}
\hatQ_{k}(\state_l, \action_l) = \bfc(\state_l, \action_l) + \gamma \min_{\action\in \actionSet}\hatQ_{k-1}(\state_l^+, \action)
\end{equation}
\STATE Estimate the function $\hatQ_k(\state,\action)$ using a regression algorithm with input pairs $(\state_l, \action_l)$ and function values $\hatQ_k(\state_l, \action_l)$.
\UNTIL{the stopping criterion is satisfied}
\STATE Compute $\hat\pi^{\ast}(\state) = \argmin\limits_{\action\in \actionSet}\hatQ_k(\state,\action)$
\end{algorithmic}
\end{algorithm}
\section{System Description and Problem Setting\label{s:prob}}
\subsection{Models}
First, we briefly describe our benchmark problem - regulation of the toggle switch system (\cite{Gardner00}). The original genetic toggle switch system consists of the \laci~and \tetr~genes mutually repressing each other (see \figref{fig:ts}). We consider a generic toggle switch model; therefore, we will use numeric references for genes and proteins, that is, gene $1$ and $2$ instead of \laci~and \tetr~genes. We will refer to the protein products of genes $1$ and $2$ as proteins $1$ and $2$, respectively. We assume that for both genes the protein concentrations are given as readouts via fluorescent markers. We also assume that the control inputs are implemented as light pulses activating a photo-sensitive promoter controlling the expression of gene $1$ (\cite{Sato02}). When this photo-sensitive promoter is activated through a light pulse the concentration of protein $1$ is increased by a small amount through the expression of gene $1$.

Basic mass-action kinetics of the toggle switch result in a high-order model, which is typically reduced to a two state model using quasi steady state approximation (\cite{Guantes06}). This can be done because most of the reactions (including the mRNA dynamics and the light-induction of the promoter) occur on a fast time scale (order of seconds) in comparison with the gene expression time scale (order of minutes or even hours). The reduced order model of the toggle switch system has two states, which are the two protein concentrations: 
\begin{equation}\label{eq:det_sys}
 \begin{aligned}
   n^1_{t+1} &=\beta_1 + \frac{c_1}{1 + (n^2_t)^{\alpha_2}} - c_2 n^1_t + b u_t& \\
   n^2_{t+1} &=\beta_2 + \frac{c_3}{1 + (n^1_t)^{\alpha_1}} - c_4 n^2_t&
 \end{aligned} 
\end{equation}
where $n^i_t$ is the concentration of protein $i$ at time $t$, $c_1$ and $c_2$ are the effective rate of synthesis of the repressors, $\alpha_i$ is the cooperativity coefficient of the repressor $i$, $c_2$ and $c_4$ are the degradation rates of proteins, $\beta_i$ models leaky transcription during the gene expression of the gene $i$, and $b$ is the increase in protein concentration produced per unit of time as a result of one light pulse. We approximate the action of light induction $u_t$ as a discrete variable in the set $U = \{0, 1\}$. A more realistic model would also have a time-delayed control action. Such an extension requires simple modifications of our control algorithm, but makes the results less transparent and harder to analyse. 
In our simulations we use a training model and a validation model.
\begin{equation}\label{eq:systems}
 \begin{aligned}
   &\textrm{ Training model}&\\
   n^1_{t+1} &=0.1+\frac{30}{1 + (n^2_t)^2 } - n^1_t + 20 u_t& \\
   n^2_{t+1} &=0.1+\frac{60}{1 + (n^1_t)^{2}} - n^2_t&
 \end{aligned}\hspace{0.1\columnwidth}  \begin{aligned}
   &\textrm{ Validation model }&\\
   n^1_{t+1} &=0.1+\frac{60}{1 + (n^2_t)^{2}} - n^1_t + 20 u_t& \\
   n^2_{t+1} &=0.1+\frac{30}{1 + (n^1_t)^2}   - n^2_t&
 \end{aligned} 
\end{equation}
Both models are bi-stable toggle switches with quantitatively different behaviours. Moreover, the steady states of the validation model are relatively far from the steady states of the training model. The stable steady states of the training model are approximately at ${\se}_1 = \begin{pmatrix} 0.11 & 29.26 \end{pmatrix}$ and ${\se}_2 = \begin{pmatrix} 59.4 & 0.17 \end{pmatrix}$ concentration units, while the stable steady states of the validation model are at ${\se}_1 = \begin{pmatrix} 59.4 & 0.17 \end{pmatrix}$ and ${\se}_2 = \begin{pmatrix} 0.11 & 29.26 \end{pmatrix}$.  Such a situation is possible in biological applications, for example, due to different cell behaviours within a population of cells. Moreover, even for a single cell, different experiments may produce values of parameters with a large variation. 
 

\subsection{Control Algorithm}
Our goal is to develop a control algorithm, which learns how to near-optimally control the toggle switch system in a single experiment. Toggling the switch can be done experimentally in a couple of hours and the fastest measurement sampling is in the order of one minute. This gives at most $200$ samples in a single trajectory. Learning a near-optimal control policy for toggling the switch with such limited amount of data is an extremely hard problem to solve. To tackle this issue, we propose to first learn a ``rough approximation'' of the control policy obtained by applying Algorithm~\ref{alg:fqi} to one-step system transitions $\calF$ artificially generated from simulations of a mathematical model of a genetic toggle switch. Afterwards the policy is fine-tuned by mixing the online measurements with past observations $\calF$. The Exploration/Exploitation trade-off is addressed using an $\varepsilon$-greedy policy. 
 
Our approach is outlined in Algorithm~\ref{alg:fqi_online}. Let $\hatQ_{\rm Alg~\ref{alg:fqi}}(\cdot,\cdot)$ be the approximation of the $Q$ function obtained by Algorithm~\ref{alg:fqi}. This will be the initial $Q$ function denoted as $\tildeQ_{\rm cur}$. Then we assume that $T_{\rm update}$  direct interactions with the real system are performed by computing actions using $\tildeQ_{\rm cur}$ and new input-output samples are collected in $\calF_{\rm new}$. Given the set $\calF_{\rm new}$, the approximation $\tildeQ_{\rm cur}$ of the $Q$ function is updated as prescribed in Algorithm~\ref{alg:fqi_online}. After that the new set $\calF_{\rm new}$ is formed and new samples are collected. 

The major challenge of Algorithm~\ref{alg:fqi_online} is appropriately choosing the function $h(\cdot,\cdot)$, which combines the sets $\calF_{\rm cur}$ and $\calF_{\rm new}$. As an example, we consider $h(\calF_{\rm cur}, \calF_{\rm new}) = \calF_{\rm cur}\cup \calF_{\rm new}$. Such a choice has some drawbacks. If the initial set $\calF$ contains many samples, then the updates in~\eqref{eq:online} will not result in significant changes in the policy. This happens because the algorithm appreciates equally the samples in $\calF$ and the new sets $\calF_{\rm new}$, even though the samples in $\calF$ are artificially generated using a mathematical model and the samples in $\calF_{\rm new}$ are obtained from the real system.


\begin{algorithm}[t]
\caption{Online learning algorithm~\label{alg:fqi_online}}
{\bf Inputs:} Set $\calF = \{\state_l, \action_l, \state_l^+\}_{l=1}^{\#\calF}$, cost function $\rew(\cdot,\cdot)$, function $\hatQ_{\rm Alg~\ref{alg:fqi}}(\cdot,\cdot)$,  number of iterations $N$, function $h (\cdot,\cdot)$
\begin{algorithmic}
\STATE $\tildeQ_{\rm cur}(\cdot,\cdot)\leftarrow \hatQ_{\rm Alg~\ref{alg:fqi}}(\cdot,\cdot)$
\STATE $\calF_{\rm cur}\leftarrow \calF$
\WHILE{new data is received}
\STATE $\pi(\state) = \min\limits_{\action' \in \actionSet}\tildeQ_{\rm cur}(\state, \action')$
\STATE $k \leftarrow 1$ , $i \leftarrow 1$ 
\WHILE{$i \le T_{\rm update}$} 
 \STATE compute $\action_i = \pi(\state_i)$ 
 \STATE observe the successor state  $\state_{i+1}$ for the state-action pair $(\state_i, \action_i)$.
 \STATE $i \leftarrow i + 1$ 
\ENDWHILE
\STATE Collect a set of new samples $\calF_{\rm new} = \{\state_m, \action_m, \state_m^+\}_{m=1}^{T_{\rm update}}$
\STATE $\calF_{\rm cur} \leftarrow h(\calF_{\rm cur}, \calF_{\rm new})$
\WHILE{ $k\le N$}
\STATE In order to obtain the values of $\tildeQ_{k+1}(\cdot, \cdot)$ for all $\{\state_l, \action_l\}$ in $\calF_{\rm cur}$ compute: 
\begin{equation}
\label{eq:online}
\tildeQ_{k+1}(\state, \action) = \rew(\state, \action) + \gamma \min\limits_{\action'\in\actionSet}\tildeQ_{k}(\state^+, \action')
\end{equation}
\STATE Estimate the function $\tildeQ_{k+1}(\state,\action)$ using a regression algorithm with input pairs $(\state_l, \action_l)$ and function values $\tildeQ_{k+1}(\state_l, \action_l)$.
\STATE $k \leftarrow k+1$ 
\ENDWHILE
\STATE $\tildeQ_{\rm cur}(\cdot,\cdot)\leftarrow \tildeQ_k(\cdot,\cdot)$
 \ENDWHILE
\end{algorithmic}
\end{algorithm}

An important task of such a learning algorithm is a trade-off between exploration and exploitation during the generation of the set $\calF_{\rm new}$. Exploration is required, since the real system is essentially unknown to the algorithm and the exploratory actions will provide new information. The trade-off policy between exploration and exploitation is defined as follows:
\[
\action_t = \begin{cases}
        \argmin_{\action'\in\actionSet} \tildeQ_k(\state_t, \action') & \textrm{ with probability $1-\varepsilon_t$} \\
        \textrm{ random action} & \textrm{ with probability $\varepsilon_t$}
       \end{cases}
\]
where $\state_t$ is the state measured at time $t$ and $\tildeQ_k(\cdot, \cdot)$ is a current approximation of the $Q$ function. In our experiment $\varepsilon_t$ is a decreasing function of $t$ between zero and one. During the first time samples, the need for new information is typically higher, and thus a high value of $\varepsilon_t$ should be chosen. 
 
\subsection{Parameters of the Algorithm \label{s:param}}
The structure of the instantaneous cost $\rew(\state, \action)$ is chosen as follows:
\[
 \rew(n^1, n^2, u) =  \max\left(n^1/\alpha_1, n^2/\alpha_2\right) - \min\left(n^1/\alpha_1, n^2/\alpha_2\right) + \alpha_u u
\]
where $\alpha_1$, $\alpha_2$, $\alpha_u$ are non-negative constants. The function
\[\max\left(n^1/\alpha_1, n^2/\alpha_2\right) - \min\left(n^1/\alpha_1, n^2/\alpha_2\right) \]
appears in studies on consensus theory as a Tsitsiklis Lyapunov function. The vector $\begin{pmatrix}
  \alpha_1 & \alpha_2 \end{pmatrix}$ can be seen as the target point of the control algorithm and the function itself can be viewed as a metric. Since only the ratio between the protein concentrations and the constants $\alpha_i$ appears in the cost, the algorithm is robust towards changes in $\alpha_i$, which are within one order of magnitude of $\alpha_i$. The major requirement is that $\alpha_1$ is much larger than $\alpha_2$, which forces the protein concentration $n^1$ to be much larger than the protein concentration $n^2$. Note that instead of a Tsitsiklis Lyapunov function other functions can be used, for example, a distance in $l_p$, a linear Lyapunov function $n^1/\alpha_1+n^2/\alpha_2$ etc. However, the main concern of this work is evaluating the performance of the online algorithm; therefore, the choice of the cost function will be addressed in future work. The term $\alpha_u u$ penalises the control signal and therefore attempts to minimise the burden associated with light-induced gene expression. The choice of $\alpha_u$ dictates the trade-off that exists between toggling the switch fast and toggling the switch with a reduced gene expression burden. We choose parameters $\alpha_u$ and $\gamma$ by tuning. The parameter $\alpha_u$ is equal to one in the simulations, and the discount factor $\gamma$ is set to $0.75$. 

Computing the control actions is a cheap procedure; however, performing the updates of the $Q$ function is a computationally harder problem. Therefore, the online algorithm performs $10$ iterations of the fitted $Q$ algorithm every $10$ time samples, in order to emulate computationally constrained controllers. The number of input-output samples used for computing the initial policy for the online algorithm is small in comparison with the purely offline algorithm. There are two reasons for such an assumption: $(a)$ it is more realistic to assume sparse input data, if we consider input-output data from previous experiments; $(b)$ fewer input-output samples imply computationally cheaper updates of the $Q$ function; $(c)$ a large amount of samples can limit the ability of the online algorithm to update the policy and the $Q$ function efficiently. 

 For the results in Figures~\ref{fig:ts_ideal} and~\ref{fig:val_offline}, we generated $1000$ trajectories with $100$ one-step transitions in each trajectory. For the simulation of the online update algorithm we generated $100$ trajectories with $100$ one-step transitions in each trajectories. The policy is updated every $10$ time samples. The stochastic simulation is performed using the direct Gillespie stochastic simulation algorithm. At every time instance $t$, one hundred trajectories starting at $\state_t$ are computed until the next time instance $t+1$, and the value $\state_{t+1}$ is then averaged over these trajectories. The average over these trajectories represents the average value of protein concentrations in a population of cells, which is much easier to measure . 

Finally, the trade-off between exploration and exploitation is decided by choosing the $\varepsilon_t$ function as follows:
\[
\varepsilon_t = \varepsilon \cdot \frac{1}{N_{\rm update} + 1}
\]
where $N_{\rm update}$ is the number of times the policy was updated online. We update the policy after $10$ time samples; therefore, $N_{\rm update}$ is equal to $O(t)$ for large $t$.
\section{Results and Discussion \label{s:res}}

\begin{figure}[t]
\centering
\incfig{
\includegraphics[width=0.45\columnwidth]{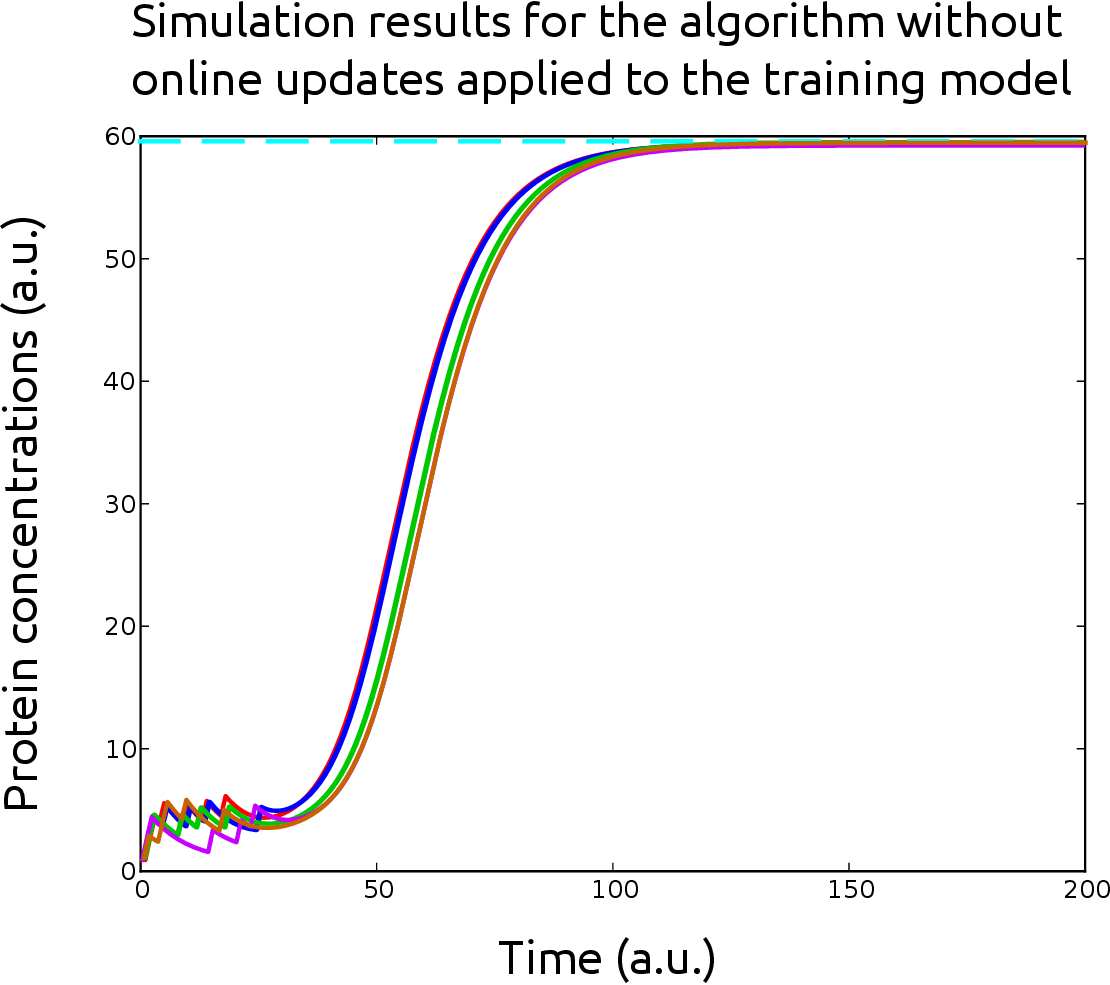}}
\caption{``Ideal'' control of the toggle switch system. The policy is computed from the input-output data of a system and then applied to control the same system. This setting is unrealistic; however, it illustrates the robustness of our control objective towards errors in the choice of the expected target point. Differently coloured trajectories correspond to the different input target points for the steady state concentration of protein $1$. We choose the target points with $15$, $30$, $45$, $60$, $90$ units, while the real upregulated steady state is approximately $60$ units (the cyan dashed line). These simulations show, that a considerable error can be made in the specification of the input target point without significant effect on the performance of the control algorithm. The schedule of light pulses is not shown due to overlapping trajectories, but all the pulses occur when the concentration of protein $1$ is smaller than $10$ units.}
\label{fig:ts_ideal}
\end{figure}
As an illustration of the benefits of the proposed approach, we investigate how it handles model uncertainty. In order to do so, we specify a training model and a validation model as in~\eqref{eq:systems}. Both models are bi-stable toggle switches with quantitatively different behaviours. Moreover, the steady states of the validation model are relatively far from the steady states of the training model. The stable steady states of the training model are approximately at ${\se}_1 = \begin{pmatrix} 0.11 & 29.26 \end{pmatrix}$ and ${\se}_2 = \begin{pmatrix} 59.4 & 0.17 \end{pmatrix}$ concentration units, while the stable steady states of the validation model are at ${\se}_1 = \begin{pmatrix} 59.4 & 0.17 \end{pmatrix}$ and ${\se}_2 = \begin{pmatrix} 0.11 & 29.26 \end{pmatrix}$. The goal is to compute a control policy (control law), which will steer the model from the stable steady state ${\se}_1$ to the stable steady state ${\se}_2$. A control policy is a binary function of a current measurement computing the current action, which is the presence or the absence of a light pulse. Due to the systems' dynamics, a larger amount of light pulses is typically required to switch from ${\se}_1$ to ${\se}_2$ in the validation model in comparison with the training model. We are going to test our online control algorithm by computing the initial control policy from the data generated by the training model and apply this policy to the validation model. One of the challenges for an efficient control algorithm is that not only the dynamics change, but also the target steady state. This setting mimics the experimental setup, when the trajectories of the model used for the policy computation (or the training model) do no match exactly the trajectories of the real system (or the validation model). 

We first consider the deterministic case and study the robustness of our algorithm towards errors in the choice of the target steady state, that is, the presumed and a priori specified value of the upregulated steady state concentration of protein $1$. Therefore, we evaluate the proximity of five trajectories obtained with different target points, while the control policies are learned from and applied to the training model. In this case, we use the control algorithm without online updates. \figref{fig:ts_ideal} depicts the obtained trajectories of the concentration of protein $1$ associated with the gene being upregulated. The red curve corresponds to the protein concentration obtained with the input target point equal to $15$ units, the green curve corresponds to the input target point of $30$ units, the blue curve to the input target point $45$, the orange curve to $60$, and the purple curve to $90$. The upregulated steady state concentration of protein $1$ is approximately equal to $60$ units (the cyan dashed line). All the curves are very close to each other and hardly distinguishable, which indicates that our algorithm is robust to some perturbations in the choice of the target point. Note that all the light pulses occur when the protein concentration is smaller than $10$ concentration units. Hence, the policy essentially defines a threshold in the concentration of protein $1$, below which light pulses are applied and above which light pulses are not necessary since the trajectories will eventually converge to the upregulated state due to the unforced system dynamics\footnote{The optimal policy is more complicated than a simple threshold; however, the approximation of the policy by a threshold provides a general idea about the shape of the control policy}. This threshold can be adjusted by modifying the parameters of the algorithm according to the control goal: faster control or smaller burden.
\begin{figure}[t]
\centering\incfig{
\subfigure[Simulation results without online updates. The red curve corresponds to the protein concentration obtained with the input target point $15$, the green curve corresponds to the input target point $30$, the blue curve to the input target point $45$.\label{fig:val_offline}]{\includegraphics[width=0.45\columnwidth]{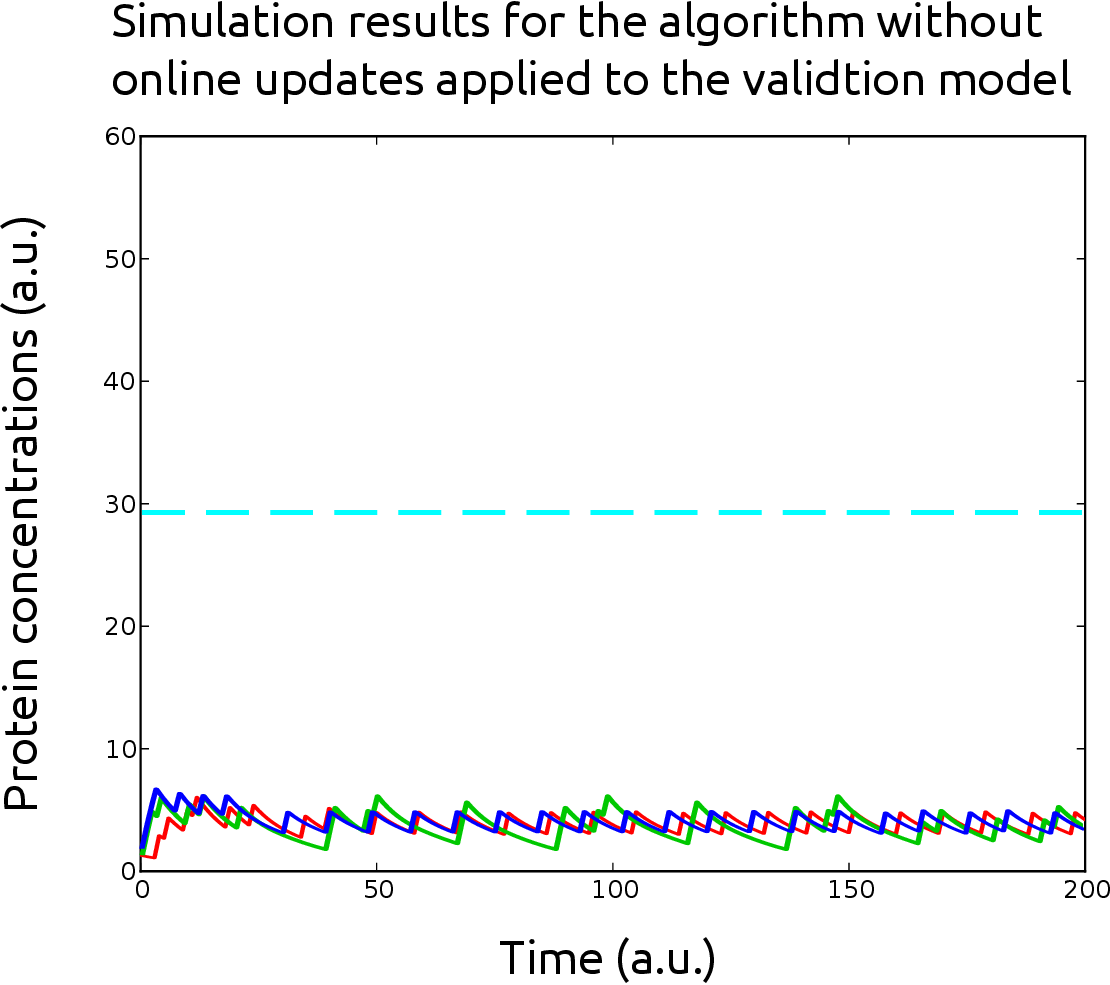}} \hspace{0.05\columnwidth}
\subfigure[Simulation results with online updates. The red curve corresponds to the simulation of our algorithm with online updates with $\varepsilon$ equal to $0.5$, the blue curve corresponds to the trajectory with $\varepsilon$ equal to $0.25$, and the green curve to $\varepsilon$ equal to $0.1$. \label{fig:val_online}]{\includegraphics[width=0.45\columnwidth]{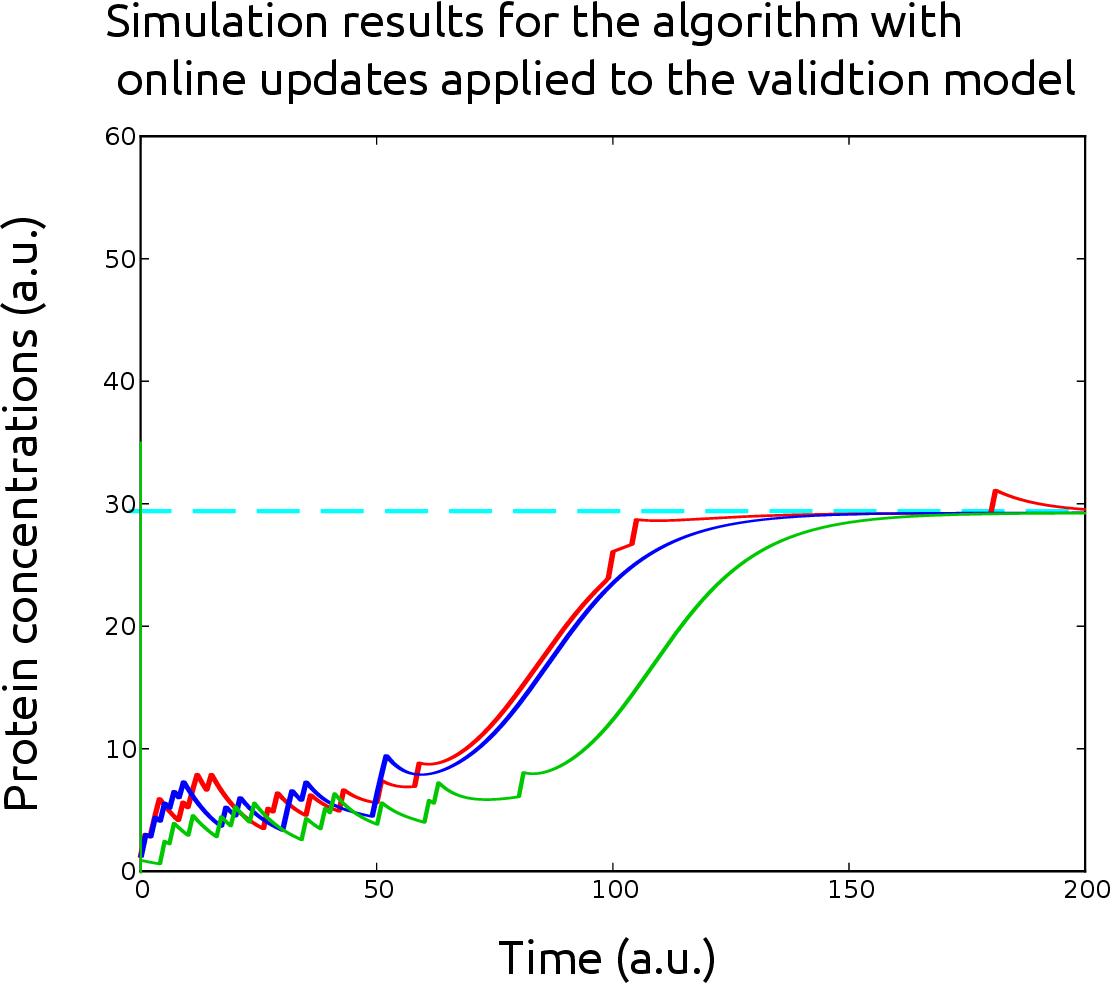}}}
\caption{Simulation results of the validation model with an initial control policy computed using the input-output data generated by the training model. In the left panel no online updates are performed, in the right panel the policy is updated using the measured input-output data. In all the simulations of the algorithm without online updates the switch is not toggled (the left panel), while in the simulations of the algorithm with online updates the switch is successfully toggled (the right panel). In both panels, the upregulated steady concentration of protein $1$ is approximately equal to $30$ (the cyan dashed line). The algorithm with online updates has two phases: exploration and exploitation. 
The trade-off between these two phases is decided by the parameter $\varepsilon$, larger values of which imply more aggressive exploration and faster learning of the system.  \label{fig:val_test}}
\end{figure}
\begin{figure}
\centering
\incfig{
\includegraphics[width=0.9\columnwidth]{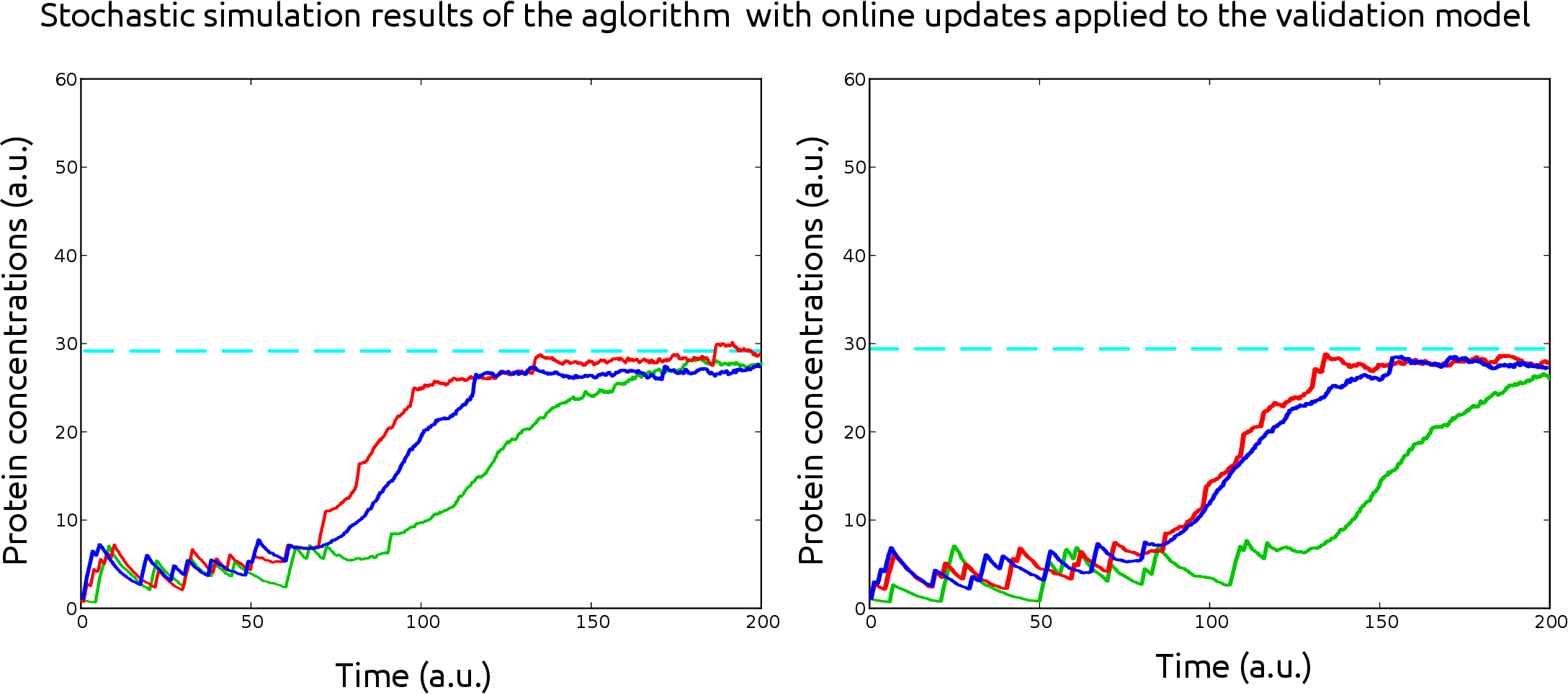}}
\caption{Stochastic simulation results of the algorithm with online updates applied to the validation model.  In the simulations in the right panel, the penalty on the amount of light pulses is twice larger than in the simulations in the left panel. The red lines correspond to the value of $\varepsilon$ equal to $0.5$, the blue lines corresponds to the value of $\varepsilon$ equal to $0.25$, and the green lines to $\varepsilon$ equal to $0.1$.} \label{fig:stoch_test}
\end{figure}

However, the setting when the policy is learned from a system and then used to control the same system is not entirely realistic. Typically, some model parameter variations are present. Here we model the case when the validation system has a considerable difference in parameter values in comparison with the training system. In \figref{fig:val_offline}, we depict the simulated trajectories in such a situation. We run the algorithm with three different target points: $15$ (the red curve), $30$ (the green curve), and $45$ (the blue curve). The actual upregulated protein concentration is approximately equal to $30$. In all the simulations the algorithm without online updates cannot force the system into the upregulated state (the cyan dashed line). This occurs because the threshold required to ensure the switch in the validation model is higher than the one computed using the training model.

In \figref{fig:val_online}, we present simulation results of the proposed algorithm with online updates. The algorithm collects new input-output samples and updates the policy at certain time intervals. The algorithm alternates between two phases: exploration and exploitation. In the exploitation phase the algorithm steers the system towards the specified goal using the control policy computed so far. In the exploration phase the algorithm generates data by randomly choosing ``to apply a light pulse'' or ``do nothing''. Due to these random choices, data generated during the exploration phase is not correlated with the past samples. 
A major challenge in this algorithm is deciding the trade-off between the exploitation and exploration phases. A simple heuristic for tackling this trade-off is as follows. At time $t$, with a probability $\varepsilon_t$ explore the system, and with a probability $1-\varepsilon_t$ exploit the system. There is a bigger need in exploration in the beginning of the experiment; therefore, $\varepsilon_t$ should be larger for small $t$ and decrease with time. Hence,  we choose $\varepsilon_t$ as $\varepsilon\cdot \sigma(t)$, where $\varepsilon$ is a positive constant smaller than one and $\sigma(t)$ is a monotonically decreasing function of $t$ such that $\varepsilon\cdot\sigma(t)$ is always larger than zero and smaller than one. Larger values of $\varepsilon$ indicate more aggressive exploration and faster learning. In these simulations, the input target point for the concentration of protein $1$ is equal to $15$. In \figref{fig:val_online}, the red curve corresponds to the simulation of our algorithm with online updates with $\varepsilon$ equal to $0.5$, the blue curve corresponds to the trajectory with $\varepsilon$ equal to $0.25$, and the green curve to $\varepsilon$ equal to $0.1$. The cyan dashed line represents the upregulated steady state concentration of protein $1$. In all the simulations the switch is successfully toggled for the validation model, even if the initial policy is obtained by learning from the training model. 

One of the biggest advantages of our approach is the ability to handle stochastic dynamics without any modifications of the algorithm. Moreover, behaviours of the controlled toggle switches in the stochastic case are qualitatively similar to the deterministic case. We present the simulation results with online updates for a similar setting as in the deterministic case in the left panel of~\figref{fig:stoch_test}. Additionally, we present the simulation results in the setting with a twice as large penalty on the amount of applied light pulses in the right panel of~\figref{fig:stoch_test}. In both figures, the red curves correspond to the simulation of our online algorithm with $\varepsilon$ equal to $0.5$, the blue curves correspond to  $\varepsilon$ equal to $0.25$, and the green curves to $\varepsilon$ equal to $0.1$. It is noticeable that toggling the switch takes longer with a larger penalty on the amount of light pulses. However, the main outcome of these simulations is that our algorithm can be applied to systems with stochastic dynamics, and as a consequence can potentially handle wet-lab data efficiently.

Our algorithm, however, does not take into account the \emph{a priori} knowledge that it is being applied to a different, but structurally similar system. A correct exploitation of structural similarity between the learned from and applied to systems may significantly improve the performance of the presented algorithm. This constitutes one of the main directions for future work that is currently under investigation.

As a final remark, we have shown that the presented framework can efficiently control a (stochastic) model of the genetic toggle switch with a parametric uncertainty. The major feature of our control algorithm is its learning nature. The algorithm computes an initial control policy using input-output data obtained from simulations of a training model, and after that updates the policy by using the input-output data obtained from the validation model (or the real system). In the presented example, despite the fact that the training and validation models had quite different quantitative behaviours the control objective was always reached using our online control method. This indicates a potential for a generalisation of this data-based control method to more complex gene regulatory networks. 

\section*{Acknowledgement}
Aivar Sootla and Guy-Bart Stan acknowledge the support of EPSRC through the project EP/J014214/1 and the EPSRC Science and Innovation Award EP/G036004/1. Damien Ernst acknowledges support of  the Belgian Network DYSCO (Dynamical Systems, Control, and Optimization), funded by the Interuniversity Attraction Poles Programme, initiated by the Belgian State, Science Policy Office.

\bibliography{Biblio}
\end{document}